\documentclass[aps]{revtex4}
\usepackage{epsf,graphicx} 
\newcommand{\bb}{\begin{eqnarray}}
\newcommand{\ee}{\end{eqnarray}}
\begin{document}
%{\begin{flushright}
%SINP/TNP/02-8\\
%TIFR/TH/02-10
%\end{flushright}
%} 
\title{\bf Thermodynamics Of dilaton-axion black holes}
\author{Tanwi Ghosh\footnote{E-mail: tanwi.ghosh@yahoo.co.in} and Soumitra SenGupta\footnote{E-mail: tpssg@iacs.res.in}}
\affiliation{Department of Theoretical Physics , Indian Association for the
Cultivation of Science,\\
Jadavpur, Calcutta - 700 032, India}
\vskip 5cm

\begin{abstract}
Considering a generalised action for Einstein Maxwell theory in four dimensions coupled to scalar and pseudo-scalar fields, 
the thermodynamic properties of asymptotically flat black holes solutions in such a background are investigated.
Bekenstein-Hawking area-entropy law is verified for these class of black holes.
From the property of specific heat, it is shown that such black holes can be stable 
for certain choice of the parameters like charge, mass and the scalar vacuum expectation value. The possibility of 
a black hole phase transition is discussed in this context.
\end{abstract}
\maketitle
%\vfill

{{\Large {\bf Introduction}}}\\
Theories of gravitation with background scalar and pseudo-scalar field have been studied extensively
for a long time. In particular in  string inspired models, the electromagnetic-dilaton coupled black hole solutions 
obtained by Garfinkle,Horowitz, Strominger initiated intense 
activities in these areas\cite{gar}. Among various properties of such black hole
solutions , the study of their thermodynamic properties has always been an important area of research in the Physics
of black holes.

String theory based-models contain two massless scalar fields, dilaton and axion,
in the low energy effective action in four dimension. The axion $\zeta$ is connected via a duality 
transformation to the three form field strength $H_{\mu\nu\lambda}$ corresponding to the two form Kalb-Ramond 
antisymmtric tensor field
$B_{\mu\nu}$ which  appears as a massless closed string mode.
While in the low energy action the dilaton and axion couple to the electromagnetic field in a specific way, a more 
generalised coupling with  Einstein and Maxwell theory in four dimensions were proposed both for  
asymptotically flat and non-flat dilaton-axion black holes. The couplings in  general depend on the vacuum expectation
values of the various moduli of compactification. Using the duality symmetry in string theory it has been shown that
starting from one particular spacetime solution 
with dilaton and axion one can generate inequivalent solutions for the spacetime as well as dilaton and axion.
Corresponding black hole solutions have also been studied extensively\cite{sha,se,se1,or,kal,horo}.
This prompted a parametric generalization of this coupling and treat them as independent
parameters to study various generic scalar coupled black hole solutions \cite{sou}. It is worthwhile therefore
to explore various thermodynamic properties of this class of black holes and verify the Bekenstein-Hawking
area-entropy conjecture in this context. Moreover   
as the parameters $a$ and $b$ give us the relative strength
of the scalar and pseudo scalar couplings, we propose to investigate in this work, how these couplings modify the
thermodynamic properties of these black holes.

The generalized Einstein-Maxwell -dilaton-axion action is given as \cite{sou};
\begin{eqnarray}
S=\int d^4x\sqrt{-g}[\frac{1}{2\kappa}(R-\frac{1}{2}\partial_{\mu}\varphi\partial^{\mu}\varphi
-\frac{1}{2}e^{2a\varphi}\partial_{\mu}\zeta\partial^{\mu}\zeta)-e^{-a\varphi}F_{\mu\nu}F^{\mu\nu}
-b\zeta F_{\mu\nu}*F^{\mu\nu}]
\end{eqnarray} 
where $a$ and $b$ are coupling parameters of dilaton field $\varphi$ and axion field $\zeta$ respectively with the Maxwell field tensor 
$F_{\mu\nu}$ and its dual $*F^{\mu\nu}$.
The corresponding black holes solutions and their thermodynamic properties are interesting areas of studies  
from the point of view of such generalized scalar coupling. 
It may be recalled that the thermodynamic properties of various black holes have been extensively investigated in 
different context\cite{davies,geor,aman,yun,jo,ghosh,
page,saurya,berezin,radu,terry,shung,wang,ju,ji,li,james,sergi,bekenstein,chamb,haw,caldarelli,mitra,lemos,louko,haw2,berman,
cvetic,hooft,bbs,smarr,bbw,wy,anm,wang1,gubsar,tanwi,tgssg}.
In this work we shall study the thermodynamic properties of the asymptotically 
flat dilaton-axion black holes  
as obtained from the above action \cite{sou} .  
After deriving the general expressions for the Hawking temperature and entropy for arbitrary value of $a$ and $b$, we shall 
consider two different cases :  $|b|=|a|$, which implies that the dilaton coupling parameter $a$ and the axion coupling
parameter $b$ with electromagnetic field are equal, 
and $|b|\neq |a|$ with $b << 1$ implying that the axion coupling is much weaker. 
Area-entropy law as well as first law of black hole mechanics are verified for these class of black holes. 
We shall further show that these two
cases yield distinct results in terms of thermodynamic properties of black holes, indicating the 
different roles of the scalar and pseudo scalars in determining the thermodynamic properties. Our result
brings out that with increase in axion coupling, the black hole may undergo second order phase transition.

\vskip 1cm

%\vfill
{{\Large {\bf Asymptotically Flat Dilaton-Axion Black hole}}}\\

Considering above mentioned Einstein-Maxwell-Kalb-Ramond action for arbitrary coupling parameters a and b, static spherically symmetric black hole solution is given by \cite{sou},
\begin{eqnarray}
ds^2= -\frac{(r-r_+)(r-r_-)}{(r-r_0)^{2-2n}(r+r_0)^{2n}}dt_s^2
+\frac{(r-r_0)^{2-2n}(r+r_0)^{2n}}{(r-r_+)(r-r_-)}dr^2+\frac{(r+r_0)^{2n}}{(r-r_0)^{(2n-2)}}
[d\theta^2+sin^2\theta d\phi^2]    
\end{eqnarray}\\
where electromagnetic field components and black hole horizons are\\

 $F_{tr}=\frac{(q_e-q_mb\zeta)e^{a\varphi}}{(r-r_0)^{2(1-n)}(r+r_0)^{2n}}dt dr$ and
 $F_{\theta\phi}=q_m sin{\theta}d\theta d\phi$\\

$r_{\pm} = m_0 \pm\sqrt{m_0^2+r_0^2-\frac{1}{8}(\frac{K_1}{n}+\frac{K_2}{1-n})}$ ; $r_0=\frac{1}{16m_o}(\frac{K_1}{n}-\frac{K_2}{1-n})$
 
$m_0= m-(2n-1)r_0$ ; $K_1 = 4n[4r_0^2 +2kr_0(r_++r_-)+k^2r_+r_-]$ ;  $K_2=4(1-n)r_+r_-$;  $0<n<1$ 
and $ m = \frac{1}{16r_0}(\frac{K_1}{n}-\frac{K_2}{1-n})+(2n-1)r_0$, where $m$ is the mass of the 
black hole and k=1 for asymptotically flat case.\\
For these class of  black hole, the surface gravity $\kappa$ can be easily obtained as,
\begin{eqnarray}
\kappa = \frac{(r_+-r_-)}{(r_0+r_+)^{2n}(r_+-r_0)^{2-2n}}
\end{eqnarray}
This gives the the Hawking temperature T of the black hole as,
\begin{eqnarray}
T=\frac{\kappa}{2\pi}=\frac{(r_+-r_-)}{4\pi(r_0+r_+)^{2n}(r_+-r_0)^{2-2n}}
\end{eqnarray}
In order to determine the entropy, the expression of area $A$ can be derived from,
\begin{eqnarray}
A=\int\sqrt{g_{\theta\theta}g_{\phi\phi}}d\theta d\phi
\end {eqnarray}
For the above metric this yields,
\begin{eqnarray}
=4\pi \frac{(r_++r_0)^{2n}}{(r_+-r_0)^{2n-2}}
\end{eqnarray}
To study further  thermodynamic behaviour as well as phase transitions for such dilaton-axion black hole, 
we now consider the following special cases in respect to dilaton and axion coupling parameters $a$ and $b$.\\

{{\large {\bf Case - I : $a = b$}}}\\
This case corresponds to the solution when the dilaton and axion couples with equal strength with the Maxwell field. 
For asymptotically flat case we begin with this condition i.e. 
$|a|=|b|$ .In  this case black hole metric solution takes the form \cite{sou};
\begin{eqnarray}
ds^2=-(1-\frac{2m_0}{r})(1-\frac{2r_0}{r})^{\frac{1-a^2}{1+a^2}}dt^2+
(1-\frac{2m_0}{r})^{-1}(1-\frac{2r_0}{r})^{\frac{a^2-1}{a^2+1}}dr^2+
r^2(1-\frac{2r_0}{r})^{\frac{2a^2}{1+a^2}}d\Omega^2
\end{eqnarray}
where parameters $m_0$,$r_0$ ,the black hole mass m  total charge Q,electric charge $Q_e$ and
magnetic charge $Q_m$ are related as:
$r_0=\frac{(1+a^2)Q^2e^{-a\varphi_0}}{4m_0}$,$m_0=m-\frac{(1-a^2)}{(1+a^2)}r_0$, $Q^2=Q_e^2+Q_m^2$ with $\varphi_0$ as the asymptotic 
value of the scalar field. These solutions represent black holes with it's horizon  located at $r=r_+=2m_0$. 
Such solution clearly has a curvature singularity at $r=2r_0$.\\

The action for this spherically symmetric dilaton-axion black hole has the well-known form \cite{li}\\

\begin{eqnarray}
I_{ac} = \frac{\beta}{2}(m-Q_e\Phi)
\end{eqnarray}
where $Q_e$ is the electric charge with the corresponding 
potential  $\Phi =\frac{Q_e e^{-a\varphi_0}}{r_+}$ .Substituting for $r_+$ and m the action becomes, 
\begin{eqnarray}
I_{ac} = \frac{\beta}{2}(m_0-\frac{Q_e^2 e^{-a\varphi_0}}{2m_0}+\frac{(1-a^2)r_0}{(1+a^2)})
\end{eqnarray}
The thermodynamic potential W for the corresponding  grand canonical ensemble is given as,
$W=E-TS-Q_e\Phi$, where S and E are the entropy and energy of the black hole.
From the expressions of the action we can find out thermodynamic quantities as follows\cite{li,caldarelli}:
\begin{eqnarray}
W=\frac{I_{ac}}{\beta}=\frac{1}{2}(m_0-\frac{Q_e^2 e^{-a\varphi_0}}{2m_0}+\frac{(1-a^2)}{(1+a^2)}r_0)
\end{eqnarray}
The potential $\Phi$ has the well known form
\begin{eqnarray}
\Phi =(\frac{\partial W}{\partial Q_e})_\beta=\frac{Q_e e^{-a\varphi_0}}{2m_0}=\frac{Q_e e^{-a\varphi_0}}{r_+}
\end{eqnarray}
Using equation (5) this becomes,
\begin{eqnarray}
\Phi =\frac{Q_e e^{-a\varphi_0}}{2m_0}=\frac{Q_e e^{-a\varphi_0}}{r_+}
\end{eqnarray}
The corresponding free energy  
\begin{eqnarray}
F=E-TS=W+Q_e\Phi=\frac{1}{2}(m_0+\frac{Q_e^2 e^{-a\varphi_0}}{2m_0}+\frac{(1-a^2)}{(1+a^2)}r_0)
\end{eqnarray}
and the surface gravity
\begin{eqnarray}
\kappa= \frac{2m_0}{r_+^2}(1-\frac{2r_0}{r_+})^{\frac{(1-a^2)}{(1+a^2)}}
\end{eqnarray}
The Hawking temperature of the black hole is now easily determined as, 
\begin{eqnarray}
T=\frac{\kappa}{2\pi}=\frac{1}{4\pi}\frac{2m_0}{r_+^2}(1-\frac{2r_0}{r_+})^{\frac{(1-a^2)}{(1+a^2)}}=\frac{1}{4\pi}\frac{1}{r_+}(1-\frac{2r_0}{r_+})^{\frac{(1-a^2)}{(1+a^2)}}
\end{eqnarray}
Now using \cite{hooft,tanwi,tgssg}
\begin{eqnarray}
S=\beta^2(\frac{\partial F}{\partial\beta})_{Q_e}
\end{eqnarray}
one obtains
\begin{eqnarray}
S = \pi r_+^2(1-\frac{2r_0}{r_+})^{\frac{2a^2}{(1+a^2)}}
\end{eqnarray}
To verify the Bekenstein-Hawking \cite{bekenstein} area-entropy  law, the area A of the black hole can now be derived as,
\begin{eqnarray}
A &&=\int\sqrt{g_{\theta\theta}g_{\phi\phi}}d\theta d\phi
\nonumber =4\pi r_+^2(1-\frac{2r_0}{r_+})^{\frac{2a^2}{(1+a^2)}}
\end{eqnarray}
Comparing with the expression of entropy above, we find that 
entropy S and area A are related by
\begin{eqnarray}
S=\frac{A}{4}
\end{eqnarray}
This confirms the Bekenstein-Hawking area-entropy law for these new class of scalar coupled black holes.
Substituting  the expressions of $r_0$ and $r_+$,one can further write the entropy as,
\begin{eqnarray}
S=\pi r_+^2[1-\frac{(1+a^2)Q_e^2e^{-a\varphi_0}}{4m_0^2}]
^{\frac{2a^2}{(1+a^2)}}
\end{eqnarray}
From the above expressions of temperature,entropy and potential we immediately obtain, 
\begin{eqnarray}
TdS +\Phi dQ_e=dm
\end{eqnarray} 
where we have substituted $\frac{Q_e^2 e^{-a\varphi_0}}{2m_0}=\frac{2r_0}{(1+a^2)}$.
This reconfirms the first law of black hole thermodynamics in this context.

For static spherically symmetric black hole,using $E=2TS +Q\Phi$,action $I_{ac}$ can be written as,
\begin{eqnarray}
I_{ac}=\frac{\beta}{2}(m_0+\frac{(1-a^2)r_0}{(1+a^2)}-\frac{Q_e^2 e^{-a\varphi_0}}{2m_0})
    =\frac{\beta}{2} m_0(1-\frac{r_0}{m_0})
    =\pi r_+^2 (1-\frac{2r_0}{r_+})^{\frac{2a^2}{(1+a^2)}}
\end{eqnarray}
which is same as the expression of the entropy. This leads to $I_{ac}=S$.
Moreover thermodynamically black hole contribution to the energy is defined as,
\begin{eqnarray}
E=(\frac{\partial(\beta F)}{\partial\beta})_{Q_e}
=m
\end{eqnarray}

We now look into the properties of the specific heat of such black holes. 
Using the well-known expression for the specific heat,
\begin{eqnarray}
C_{Q}=T(\frac{\partial S}{\partial T})_{Q}
\end{eqnarray}
the specific heat in this case becomes,
\begin{eqnarray}
\nonumber C_{Q}&&=T[2\pi r_+(1-\frac{2r_0}{r_+})^{\frac{2a^2}{(1+a^2)}}\\
&&+\frac{4a^2\pi 2r_0}{(a^2+1)}(1-\frac{2r_0}{r_+})
\nonumber^{\frac{(a^2-1)}{(a^2+1)}}][\frac{-1}{4\pi r_+^2}(1-\frac{2r_0}{r_+})^{\frac{(-a^2+1)}{(a^2+1)}}+
\frac{1}{4\pi r_+}\frac{2(-a^2+1)}{(a^2+1)}\frac{2r_0}{r_+^2}(1-\frac{2r_0}{r_+})
^{\frac{-2a^2}{(1+a^2)}}]^{-1}\\
&&=-2\pi r_+^2 (1-\frac{2r_0}{r_+})^{\frac{2a^2}{(1+a^2)}}[1-\frac{(1-a^2)Q_e^2e^{-a\varphi_0}}{4m_0^2}]
[1-\frac{(3-a^2)Q_e^2e^{-a\varphi_0}}{4m_0^2}]^{-1}
\end{eqnarray}
We now analyse the above expression of specific heat for different regime in the parameter space.
From the expression of entropy it follows 
that the entropy will be non-negative if $\frac{(1+a^2)Q_e^2e^{-a\varphi_0}}{4m_0^2}<1$.
This in turn implies that $\frac{(1-a^2)Q_e^2e^{-a\varphi_0}}{4m_0^2}<1$.
Therefore  
for $\frac{(3-a^2)Q_e^2e^{-a\varphi_0}}{4m_0^2}<1$ , specific 
heat is negative. Such black holes therefore  can not be 
stable locally. \\
However for $\frac{(3-a^2)Q_e^2e^{-a\varphi_0}}{4m_0^2}>1$ 
along with $\frac{(a^2+1)Q_e^2e^{-a\varphi_0}}{4m_0^2}<1$ ,one finds that $a^2<1$ and in such a scenario
the specific heat clearly becomes positive and the  black hole is stable.\\ 
Moreover for  
\begin{eqnarray}
\frac{(3-a^2)Q_e^2e^{-a\varphi_0}}{4m_0^2} = 1
\end{eqnarray}
the specific heat blows up while the 
temperature and the entropy continue to be finite. This signals  a second order phase transition for such
black holes. The phase transition therefore occurs when the charge to mass ratio of the black hole 
is related to the coupling parameter $a$ and the scalar vacuum expectation value.\\ 
It is interesting to observe that when $\frac{(a^2+1)Q_e^2e^{-a\varphi_0}}{4m_0^2}=1$,
the temperature,entropy as well as the specific heat of the black hole becomes zero.

It may further be noted that for $r_0=0$ ,the expression of specific heat 
reduces to that for a Schwarzschild black hole as is expected.\\

%{{\Large {\bf Asymptotically Flat Dilaton-Axion Black hole }}}\\

{{\large {\bf Case - II : $a \neq b$}}}\\

Extending our discussion for the asymptotically flat case, we now consider a different region in the
parameter space of $a$ and $b$ for which $|a|\neq |b|$ with $a =1$ and $b<<1$.
This implies that the axion coupling charge to the Maxwell field is much smaller  than the dilaton coupling charge. 
The corresponding metric is given as,\cite{sou},
\begin{eqnarray}
ds^2=-\frac{(r-r_+)(r-r_-)}{r^2-r_0^2}dt^2+\frac{r^2-r_0^2}{(r-r_+)(r-r_-)}dr^2+(r^2-r_0^2)d\Omega^2
\end{eqnarray}
where $r_0=\frac{(Q_e^2-Q_m^2)e^{-\phi_0}}{2m}$ and event horizons are located at $r_\pm =m 
\pm\sqrt{m^2+r_0^2-(Q_e^2+Q_m^2)e^{-\phi_0}}$. 
In this case the action can be written as,
\begin{eqnarray}
I_{ac} = \frac{\beta}{2}(m-Q_e\Phi)
\end{eqnarray}
where potential $\Phi =\frac{Q_e e^{-\phi_0}}{(r_++r_0)}$,$r_0=\frac{Q_e^2 e^{-\phi_0}}{2m}$.\\

The corresponding grand canonical potential W can be obtained from the action as,
\begin{eqnarray}
W =\frac{I_{ac}}{\beta}=\frac{1}{2}(m-\frac{Q_e^2 e^{-\phi_0}}{2m})
\end{eqnarray}
The free energy function F turns out to be, 
\begin{eqnarray}
F=E-TS=W+Q_e\Phi=\frac{1}{2}(m+\frac{Q_e^2 e^{-\phi_0}}{2m})
\end{eqnarray}
From these the expression of entropy for this black hole can be obtained as
\begin{eqnarray}
S=\beta^2(\frac{\partial F}{\partial\beta})_{Q_e}
 =\pi (r_+^2-r_0^2)
\end{eqnarray}
Calculation of horizon area $A$ in this case yields,
\begin{eqnarray}
A=\int\sqrt{g_{\theta\theta}g_{\varphi\varphi}}d\theta d\varphi
 =4\pi (r_+^2-r_0^2) 
\end{eqnarray}
Thus once again equations(24)and (25) confirms the Bekenstein-Hawking \cite{bekenstein} area-entropy law for black hole mechanics.
Substituting the expressions of $r_+$ and $r_0$, the expression for entropy reduces to,
\begin{eqnarray}
S= \pi 4m^2(1-\frac{Q_e^2e^{-\varphi_0}}{2m^2})
\end{eqnarray}
Calculating the surface gravity,$\kappa$ one finds,
\begin{eqnarray}
\kappa = \frac{(r_+-r_-)}{(r_+^2-r_0^2)}
\end{eqnarray}
From this the temperature of the black hole is found to be, 
\begin{eqnarray}
T=\frac{\kappa}{2\pi}=\frac{1}{4\pi}\frac{(r_+-r_-)}{(r_+^2-r_0^2)}
\end{eqnarray}
Combining all these together one arrives at the  first law of black hole mechanics,
\begin{eqnarray}
TdS +\Phi dQ_e=dm         
\end{eqnarray}
Similar to the Case I, energy of the black hole is E=m.\\
Substituting the expression of $\beta$, m and $\Phi Q_e$, once again the action can be written in terms of entropy S as
\begin{eqnarray}
I_{ac} = \pi(r_+^2-r_0^2)=S
\end{eqnarray}
Finally using the expression for specific heat as mentioned in CaseI, we find
\begin{eqnarray}
C_{Q}&&=\frac{-1}{8\pi m}(8\pi m )(16\pi m^2)
=-4\pi(r_++r_0)^2
\end{eqnarray}
The specific heat $C_{J,Q}$ for such black hole is therefore always negative.
So this kind of black hole is never stable locally. Once again it is easy to see that for $r_0=0$ 
the expression for $C_{J,Q}$ reduces to that for a Schwarzschild black hole.\\
It is important to observe that in Case-II, we have considered the axion coupling parameter $b$
to be very weak and much smaller than the dilaton coupling parameter $a$. Unlike Case- I, the black hole
is never stable and no phase transition occurs in this case, at least classically. This establishes an interesting
role of the pseudo scalar axion in determining the thermodynamic properties of black holes.  Our result reveals
that whenever the pseudo sclar coupling becomes as strong as the scalar coupling, the black hole can indeed 
go through second order phase transition and can therefore achieve local stability.\\    

{{\Large {\bf Conclusion}}}\\
Various thermodynamic properties for a dilaton-axion coupled black hole solutions are determined for a more general
class of their electromagnetic 
coupling parameters $a$ and $b$. It is found that for different regions in the coupling parameter space, the 
thermodynamic properties are distinct. 
The Bekenstein-Hawking area-entropy law as well as the first law of black hole thermodynamics
have been verified for these class
of scalar coupled black holes. This once again ensures the generality of the area-entropy law.
 
The issue of stability of such black holes have been examined. Analysing
the properties of the specific heat it is shown that
while for certain region in the coupling parameter space the black hole can be both stable and unstable depending on the value of it's
mass to charge ratio and the scalar vacuum expectation value, for a different choice of couplings 
they are unstable irrespective of the value of it's charge or mass. Furthermore it has been shown that a
certain relation between the mass-charge ratio and the scalar vacuum expectation value 
signals the occurrence of a second order phase transition.
The constraint on the scalar coupling parameter $a$ in such situation is determined. It is shown that in the region
where axion coupling parameter is much smaller than the dilaton coupling parameter  one find the specific heat of the black hole
to be always negative. Such unstable black hole can become stable through a phase transition when the
axion coupling parameter becomes larger and comparable to the dilaton coupling parameter . This clearly brings out 
the effects of scalar and pseudo scalar couplings on the thermodynamic properties of black holes. Our result reveals 
that even at the classical level, an appropriate choice of the axion coupling can result into a second order phase
transition.


\begin{thebibliography}{99}
\bibitem{gar} D.Garfinkle,G.T.Horowitz and A.Strominger,Phys.Rev.D{\bf 43},3140,1991.
\bibitem{sha}A.D.Shapere,S.Trivedi and F.Wilczek,Mod.Phys.Lett.A{\bf 6},2677,1991.
\bibitem{se}A.Sen,Nucl.Phys.B{\bf 404},109,1993.
\bibitem{se1}A.Sen,Phys.Rev.Lett{\bf 69},1006,1992.
\bibitem{or}T.Ortin,Phys.Rev.D{\bf 47},3136,1993.
\bibitem{kal}R.Kallosh and T.Ortin,Phys.Rev.D{\bf 48},742,1993.
\bibitem{horo}G.T.Horrowitz and A.Strominger,Nucl.Phys.B{\bf 360},197,1991.
\bibitem{sou} S.Sur, S.Das and S.Sengupta,JHEP 0510:064,2005. 
\bibitem{davies}P.C.W.Davies,Proc.Roy.Soc.Lond{\bf A353},499,1977.
\bibitem{geor} G.Ruppeiner,Phys.Rev.D{\bf 75},024037,2007.
\bibitem{aman}Jan E.Aman,Narit Pidokrajt,Phys.Rev.D{\bf 73},024017,2006.
\bibitem{yun}Yun Soo Myung,Phys.Rev.D{\bf 77},104007,2008,Yun Soo Myung,Yong-wan Kim,Young-Jai Park,arxiv:0708.3145.
\bibitem{jo}Jose L Alvarez,F.Quevedo and A.Sanchez,Phys.Rev.D{\bf 77},084004,2008.
\bibitem{ghosh}T.Ghosh,JCAP 0411:003,2004.
\bibitem{page}Don N Page,New J Phys.{\bf 7},203,2005,Class.Quant.Grav{\bf 6},1909,1989.
\bibitem{saurya}S.Das,Pramana{\bf 63},797,2004.
\bibitem{berezin}V. Berezin,Nucl.Phys.B{\bf 661},409,2003.
\bibitem{radu}E. Radu,Mod.Phys.Lett A{\bf 17},2277,2002.
\bibitem{terry}T. Pilling,Phys.Lett B{\bf 660},402,2008.
\bibitem{shung}Shuang-Quing Wu,Phys.Lett B{\bf 608},251,2005.
\bibitem{wang}B.B.Wang and C.G.Huang,Class.Quant.Grav{\bf 19},2491,2002.
\bibitem{ju}Ju-Hua Chen,Ji-Liang Jing and Yong-Jiu Wang,Chin.Phys.{\bf 10},467,2001.
\bibitem{ji}Ji-liang Jing,Chin.Phys.Lett{\bf 14},81,1997.
\bibitem{li}Ji-liang Jing,Nucl.Phys.B{\bf 476},548,1996.
\bibitem{james}James W. York,Jr,Phys.Rev.D{\bf 33},2092,1986.
\bibitem{sergi}Sergey.N Solodukhin,Phys.Rev.D{\bf 54},3900,1996.
\bibitem{bekenstein}J.D.Bekenstein,Phys.Rev.D{\bf 12},3077,1975.
\bibitem{chamb}Chamblin,Emparan,Johnson and Myers,Phys.Rev.D{\bf 60},104026,1999,
Phys.Rev.D{\bf 60},064018,1999.
\bibitem{haw}Hawking and Page,Commun.Math.Phys{\bf 87},577,1983.
\bibitem{caldarelli}Caldarelli,Cognola and Klemm,Class.Quant.Grav{\bf 17},399,2000.
\bibitem{mitra}P.Mitra,Phys.Lett.B{\bf 441},89,1998,Phys.Lett.B{\bf 459},119,1999.
\bibitem{lemos}Peca and Lemos,Phys.Rev.D{\bf 59},124007,1999.
\bibitem{louko}Louko and Winters-Hilt,Phys.Rev.D{\bf 54},2647,1999.
\bibitem{haw2}Hawking,"Stability in Ads and Phase Transitions",talk given at Strings 99,Potsdam (Germany),july 19-24,1999.
\bibitem{berman}Berman and Parikh,Phys.Lett.B{\bf 463},168,1999.
\bibitem{cvetic}M.Cvetic and S.S.Gubsar,JHEP{\bf 07},010,1999.
\bibitem{hooft}G't Hooft,Nucl.Phys.B{\bf 256},727,1985.
\bibitem{bbs}J.M.Bardeen,B.Carter and S.W Hawking,Commun.Math.Phys{\bf 31},61,1973.
\bibitem{smarr}L.Smarr,Phys.Rev.Lett{\bf 30},71,1973.
\bibitem{bbw}H.W.Braden,J.D.Brown,B.F. Whiting and J.W.Yory,Jr,Phys.Rev.D{\bf 42},3376,1990.
\bibitem{wy}B.F.Whiting and J.W.York,Jr,Phys.Rev.Lett{\bf 61},1336,1988.
\bibitem{anm}A.Sheykhi,N Riazi and M.H.Mahzoon,Phys.Rev.D{\bf 74},044025,2006.
\bibitem{wang1}Tower Wang,Nucl.Phys.B{\bf 756},86,2006.
\bibitem{gubsar}S.S.Gubser, Nucl.Phys.B{\bf 551},667,1999.
\bibitem{tanwi} T.Ghosh, Mod.Phys.Lett A {\bf 22}, 2865, 2007
\bibitem{tgssg} T.Ghosh, S.SenGupta,Phys.Rev.D{\bf 78},024045,2008.
\end {thebibliography}
\end{document}